\documentclass[runningheads]{llncs}
\usepackage[T1]{fontenc}
\usepackage{graphicx}
\usepackage[table]{xcolor}
\usepackage{booktabs}
\usepackage{multirow}
\usepackage{makecell}
\usepackage{adjustbox}
\usepackage{amsmath}
\usepackage{float}
\usepackage{tcolorbox}
\usepackage{pbox}
\usepackage{arydshln} 
\usepackage{subcaption}
\usepackage{orcidlink}

\definecolor{Gray}{gray}{0.85}
\definecolor{LightCyan}{rgb}{0.88,1,1}

\usepackage[numbers,sort&compress]{natbib}
\definecolor{lightgray}{gray}{0.9} 
\newcommand{\lc}{\ensuremath{^\dagger}}

\tcbset{
  myprompt/.style={
    colback=gray!5,       
    colframe=gray!50,      
    fonttitle=\bfseries\scriptsize,   
    coltitle=black,        
    boxrule=0.5mm,         
    width=\textwidth,      
    top=3pt, bottom=3pt,   
    left=3pt, right=3pt,
    fontupper=\scriptsize
  }
}

\newcommand{\revised}[1]{{#1}}

\newcommand{\B}[1]{\textbf{#1}}
\newcommand{\U}[1]{\underline{#1}}

\begin{document}

\widowpenalty1
\clubpenalty1

\title{LANCER: LLM Reranking for Nugget Coverage}
%
%
\author{
Jia-Huei Ju\inst{1}\orcidlink{0000-0003-2247-3370} \and
François G. Landry\inst{2}\orcidlink{0009-0009-8412-6439} \and
Eugene Yang\inst{3}\orcidlink{0000-0002-0051-1535} \and \\
Suzan Verberne\inst{4}\orcidlink{0000-0002-9609-9505} \and
Andrew Yates\inst{3}\orcidlink{0000-0002-5970-880X}
}
\authorrunning{Ju. et al.}
%
\institute{
    University of Amsterdam, The Netherlands \\
    \email{j.ju@uva.nl} \and
    Université de Moncton, Canada \\
    \email{efl7126@umoncton.ca} \and
    Johns Hopkins University, USA \\
    \email{\{eugene.yang,andrew.yates\}@jhu.edu} \and
    Leiden University, The Netherlands  \\
    \email{s.verberne@liacs.leidenuniv.nl}
}

\maketitle              

\begin{abstract}
Unlike short-form retrieval-augmented generation (RAG), such as factoid question answering, long-form RAG requires retrieval to provide documents covering a wide range of relevant information.
Automated report generation exemplifies this setting: it requires not only relevant information but also a more elaborate response with comprehensive information.
Yet, existing retrieval methods are primarily optimized for relevance ranking rather than information coverage.
To address this limitation, we propose LANCER,\footnote{https://github.com/DylanJoo/LANCER} an LLM-based reranking method for nugget coverage.
LANCER predicts what sub-questions should be answered to satisfy an information need, predicts which documents answer these sub-questions, and reranks documents in order to provide a ranked list covering as many information nuggets as possible.
Our empirical results show that LANCER enhances the quality of retrieval as measured by nugget coverage metrics, achieving higher $\alpha$-nDCG and information coverage than other LLM-based reranking methods. 
Our oracle analysis further reveals that sub-question generation plays an essential role.
\end{abstract}

\section{Introduction}
Long-form RAG has introduced a new frontier for information-seeking.
Compared to the traditional search paradigm, users can now ask LLMs to organize information from retrieved documents. 
\revised{In this specialized generation setting, retrieval becomes crucial, because it determines the finite scope of information available for the generator to incorporate~\cite{Stelmakh2022-kt}.}
For instance, the TREC NeuCLIR track's report generation task involves open-ended, multi-faceted information needs that are addressed by generating a comprehensive report describing and citing relevant information in the corpus~\cite{Dawn2025-sg}.
This is a challenging task that requires the retrieval component to retrieve documents covering all facets of the information need, so that they can be cited in the report~\cite{Gao2023-kt}.

With these emerging use cases, there has been renewed interest in nugget-based evaluation approaches that consider what fine-grained information is provided by documents rather than considering only document-level relevance~\cite{Dietz2024-iq, Pavlu2012-ch, Pradeep2024-fe,Voorhees2004-bg}.
These approaches naturally align with the notion of information coverage~\cite{Grusky2018-ss, Over2004-yb}: \emph{how well does a retrieved context cover the required relevant facts?} Therefore, nugget coverage has become a key retrieval-sensitive criterion in long-form report generation tasks~\cite{Mayfield2024-vw}. 
In practice, however, the retrieved context often includes irrelevant and redundant information, limiting the information that the generator can use and wasting some of the generator's limited input context~\cite{Ju2025-pn}.

\revised{Existing retrieval approaches are not particularly designed for coverage~\cite{Ju2025-pn}. Neural retrieval and reranking models are typically trained to predict relevance rather than to consider nugget coverage of the retrieved context.}
While listwise rerankers can consider interactions between documents in principle, this direction is underexplored and all state-of-the-art listwise rerankers are optimized for relevance ranking (i.e., they are trained to find relevant documents, not to find a set of documents that covers all aspects of an information need)~\cite{Pradeep2023-ku, Gangi-Reddy2024-vk}.
Other approaches like bi-encoders and pointwise rerankers predict the relevance of each document independently, preventing them from considering nugget coverage across a set of documents.
Optimizing nugget coverage is closely related to diversification, which has been studied in the past, but is not the goal of any state-of-the-art ranking methods.
For example, ranking for diversification was explored using pre-neural methods~\cite{Carbonell1998-gf, diversification}, whereas generating query intents for diversification was considered with early transformer methods~\cite{macavaney2021intent5}.
Motivated by these limitations, we propose a reranking method aimed at improving nugget coverage and explore its performance on collections with fine-grained nugget judgments.

\begin{figure}[t]
    \centering
    \includegraphics[width=.95\linewidth]{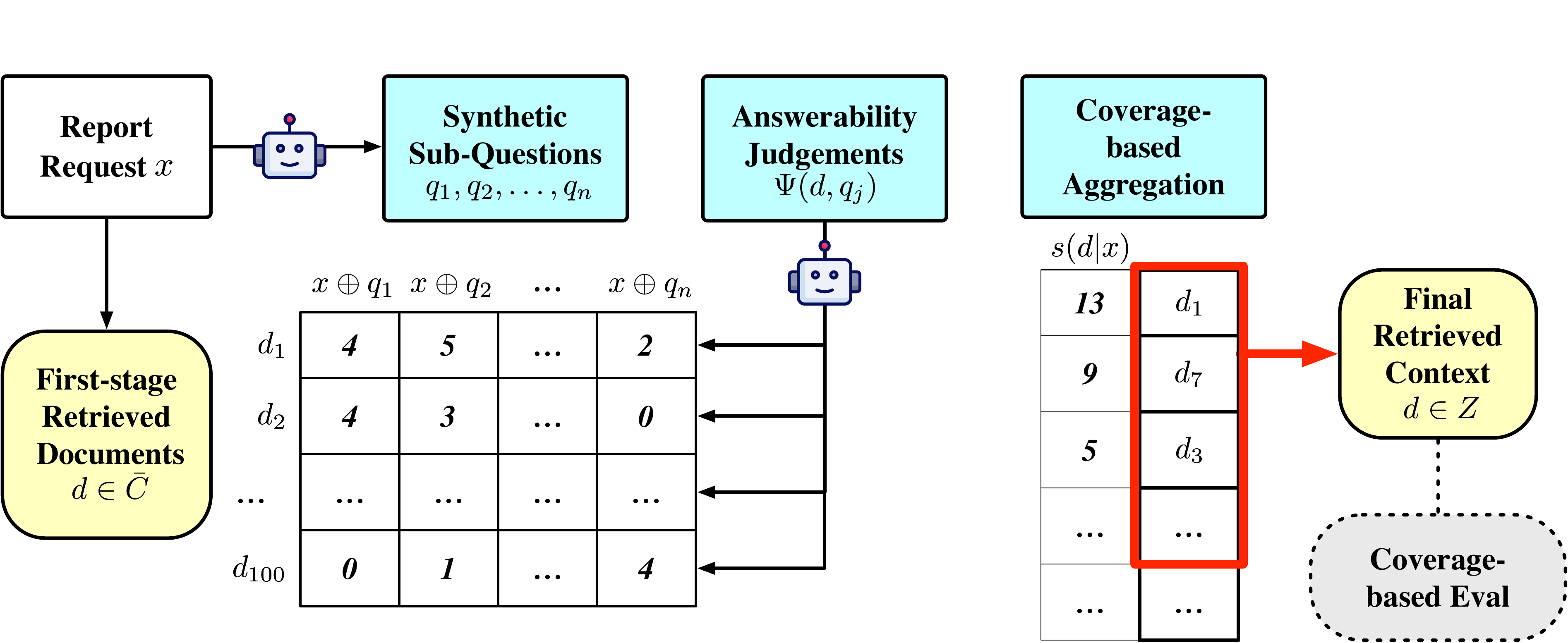}
    \caption{
    LANCER consists of three stages (blue boxes). The final retrieved context $Z$ is evaluated with nugget coverage metrics.
    }
    \label{fig:demo}
\end{figure}

We introduce LANCER, an \B{L}LM rer\B{A}nking method for \B{N}ugget \B{C}ov\B{ER}age, which aims to rerank documents in order to improve their nugget coverage at a shallow cutoff.
As illustrated in Figure~\ref{fig:demo}, LANCER has three stages:
(i) synthetic sub-question generation, 
(ii) document answerability judgment, and 
(iii) coverage-based aggregation.
LANCER uses an LLM to generate sub-questions that should be answered in order to satisfy an information need, predicts whether the documents from first-stage retrieval answer these sub-questions, and then uses these predictions to produce a reranked list that aims to cover as many information nuggets as possible. 

In our empirical evaluation on two datasets with nugget-level judgments, LANCER improves the coverage of the retrieved documents and can outperform other LLM-based reranking methods optimized for relevance~\cite{Zhuang2024-sx, Sun2023-ln}.
\revised{Moreover, LANCER offers the advantage of transparency; the synthetic sub-questions and their answerability scores provide an explicit trace of what facets of information have been collected or missed.}
\revised{In addition, providing LANCER with oracle sub-questions substantially increases performance further, demonstrating that optimizing for coverage can yield significant benefits and highlighting the quality of sub-questions as one of the important areas to improve in the future.}
We also study the impact of the parameters under different settings, providing insights into the sub-question generation and coverage-based aggregation strategies.

\section{Related Work}
Initial RAG studies~\cite{Lewis2020-ky, Guu2020-tg} have shown that retrieval can supply relevant information as a source of complementary knowledge for language models~\cite{Shi2023-kg}.
Subsequent works have further applied it on a wide range of real-world applications, e.g.~\cite{Stelmakh2022-kt, Kwiatkowski2019-uw}. 
Among them, automated report generation has unique demands for retrieval: 
it requires the retrieved context to be not only relevant but to comprehensively identify relevant documents, so the generated report can provide all relevant information in the corpus. This distinction diverges from the traditional relevance-based retrieval for short-form QA tasks, where information needs are clear and narrow.

To support the development of long-form RAG systems, many recent studies have revisited nugget-based evaluation~\cite{Pavlu2012-ch}.
A nugget represents a standalone fact, which was first introduced for evaluating definition question answering~\cite{Voorhees2003-ul} with nugget recall being the primary metric~\cite{Voorhees2004-bg}. 
The concept has been further extended to measure coverage in summarization tasks~\cite{Grusky2018-ss, Over2004-yb, Fabbri2019-hc}. 
Together, nugget and coverage collectively align with the goal of long-form RAG report generation~\cite{Mayfield2024-vw}, imposing additional coverage-based criteria on retrieval and the generated report~\cite{Dawn2025-sg, Ju2025-pn}.

However, most existing first-stage retrieval methods, instead of optimizing for coverage, are optimized solely for document relevance~\cite{Thakur2021-cc}, favoring relevant documents with common nuggets~\cite{Ju2025-pn}.
As zero-shot re-rankers, LLMs have shown their adaptability across different ranking paradigms, including pointwise~\cite{Nogueira2020-vn, Sachan2022-al}, pairwise~\cite{Qin2024-op}, listwise~\cite{Sun2023-ln, Ma2023-gb}, and setwise~\cite{Zhuang2024-sx}.
Yet each has its own drawbacks. Pointwise treats documents independently and omits relationships among redundant documents. 
While the others often focus on the relevance aspect, lacking consideration of covering more nuggets for the downstream generation.

Though coverage-based retrieval methods remain underexplored, in a similar vein, many studies have proposed to diversify the retrieved results~\cite{Chen2024-kz, Gao2024-ax, diversification, soboroff2005novelty}, aiming at tackling the trade-off between diversity and relevance~\cite{Carbonell1998-gf, Clarke2008-cn}.
Recent studies use LLMs to generate sub-queries~\cite{Li2024-aa, Zhong2025-wy} for increasing recall or intents~\cite{macavaney2021intent5} for diversification.
The research most closely related to ours is done by~\citet{Guo2024-zk}, which improves pointwise reranking with multiple criteria. Our work complements them by explicitly identifying nuggets and optimizing coverage for long-form RAG.

\section{Preliminaries} \label{sec:pre}
In this work, we aim at improving the retrieval module of an automated report generation system~\cite{Mayfield2024-vw}, which has two particular characteristics that differ from typical long-form RAG problems: 
(i) the input is a nuanced report request with multiple information needs, and
(ii) the expected output consists of sentences with citations that provide a comprehensive overview of relevant information found in a document corpus $C$. 
Formally, given a report request $x$, we define the entire report generation process as:
\begin{equation}
    y = \mathcal{G}(x, Z), \quad
    \text{where } Z \gets \mathcal{R}(x, \bar{C}). 
    \label{eq:report-gen}
\end{equation}
\revised{$\mathcal{G}$ is a report generator that takes the retrieved context $Z$ as an input for synthesizing the final report $y$.}
We define $Z$ as the retrieved context, which is the intermediate output from retrieval component $\mathcal{R}$.
Notably, we adopt the two-stage retrieval pipeline and focus on the second-stage reranking as mentioned earlier. $\bar{C}$ denotes the top-$k$ document candidates ($k \ll |C|$) retrieved from a given corpus $C$ .

To evaluate the retrieval component $\mathcal{R}$ in RAG, we assess both the intermediate retrieved context $Z$ and the final generated report $y$, representing the direct and the propagated impact of the retrieval pipeline~\cite{Es2023-yk, Ju2025-pn}. Detailed evaluation setting is depicted in Figure~\ref{fig:demo} and Section~\ref{sec:exp-setup}.

\section{Method: LANCER}
\revised{Inspired by the CRUX framework for automatically judging the information coverage of retrieved documents~\cite{Ju2025-pn}, we adapt CRUX's steps to perform reranking by removing its usage of evaluation.}
Doing so yields LANCER: an LLM reranking approach for nugget coverage optimization, which aims to increase the number of nuggets of relevant information covered.
As depicted in Figure~\ref{fig:demo}, LANCER consists of three stages:
1) \emph{generating synthetic sub-questions} that should be answered, 
2) \emph{generating sub-question answerability judgments} to predict to what extent the sub-questions are answered by documents, and
3) \emph{performing coverage-based aggregation} to rerank documents for coverage.

\subsubsection{Synthetic Sub-question Generation.}
\begin{figure}[t]
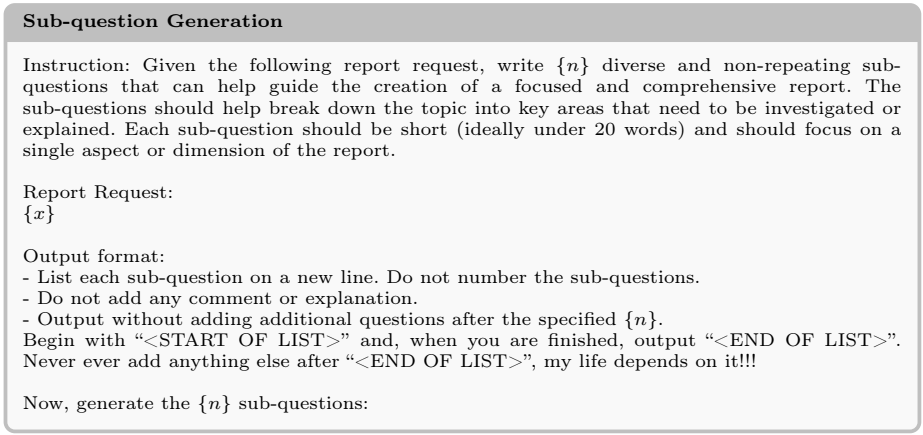

\begin{tcolorbox}[title=Sub-question Generation, myprompt]
Instruction: Given the following report request, write \{$n$\} diverse and non-repeating sub-questions that can help guide the creation of a focused and comprehensive report. The sub-questions should help break down the topic into key areas that need to be investigated or explained. Each sub-question should be short (ideally under 20 words) and should focus on a single aspect or dimension of the report. \\\\
Report Request: \\
\{$x$\} \\\\
Output format: \\
- List each sub-question on a new line. Do not number the sub-questions. \\
- Do not add any comment or explanation.  \\
- Output without adding additional questions after the specified \{$n$\}. \\
Begin with ``<START OF LIST>'' and, when you are finished, output ``<END OF LIST>''. Never ever add anything else after ``<END OF LIST>'', my life depends on it!!! \\\\
Now, generate the \{$n$\} sub-questions:
\end{tcolorbox}
\caption{Sub-question generation prompt to produce a list of sub-questions.}
\label{fig:subquestion-gen}
\end{figure}

Given a report request $x$, we first derive multiple detailed information needs by generating diverse sub-questions from the request.
We instruct an LLM to generate a set of $n$ questions that are beneficial for the downstream report generation task, denoted as $\{q_j\}_{j=1}^n$. 
The prompt we used is shown in Figure~\ref{fig:subquestion-gen}, where the report request is the only input. Detailed analysis is reported in Section~\ref{sec:ablation-nq}.

\subsubsection{Answerability Judgments Generation.}
Once the $n$ sub-questions are generated, we use the LLM to judge whether documents answer each sub-question.
Specifically, we instruct an LLM to judge the answerability of a document $d$ given a report request $x$ concatenated with the generated sub-question $q_j$:
\begin{equation}
    r_{d, q_j} = \Psi(d, x\oplus q_j),\quad\text{where } r\in [0, 5].
\end{equation}
The function $\Psi$ indicates the rubric-based LLM document judgment~\cite{Dietz2024-iq, Ju2025-pn}, which produces a rating between scale 0 and 5 using the prompt shown in Figure~\ref{fig:judge}.
Each output rating indicates the answerability of a synthetic sub-question, which collectively indicate how much each document satisfies the multi-aspect information needs of the report request $x$. 
The multi-aspect ratings are then used to rerank documents in the next stage.

\begin{figure}[t]
\begin{tcolorbox}[title=Document Answerability Judgment~\cite{Ju2025-pn, Dietz2024-iq}, myprompt]
Instruction: Determine whether the question can be answered based on the provided context? Rate the context with on a scale from 0 to 5 according to the guideline below. Do not write anything except the rating. \\\\
Guideline: \\
5: The context is highly relevant, complete, and accurate. \\
4: The context is mostly relevant and complete but may have minor gaps or inaccuracies.\\
3: The context is partially relevant and complete, with noticeable gaps or inaccuracies.\\
2: The context has limited relevance and completeness, with significant gaps or inaccuracies.\\
1: The context is minimally relevant or complete, with substantial shortcomings.\\
0: The context is not relevant or complete at all. \\\\
Question: \{$q$\} \\ Context: \{$c$\} \\ Rating:
\end{tcolorbox}
\caption{Rubric-based answerability judgment prompt. The output rating is converted into 0 to 5, and the output with incorrect formats is assigned to 0.}
\label{fig:judge}
\end{figure}

\subsubsection{Coverage-based Aggregation Strategies.} \label{sec:strategy}
In this step, we use the multi-aspect ratings to produce a reranked list that optimizes for coverage.
To do so, we explore several coverage-based aggregation strategies, including simple summation, rank fusion, and greedy selection. 

\paragraph{Summation (sum \& sum-$\tau$).}
A straightforward strategy is to sum $n$ ratings to produce a single score for each document $d$:
\begin{equation}
    s_{sum}(d| x) = \sum_{j=1}^{n} r_{d, q_j}. \label{eq:sum}
\end{equation}
In addition, we experiment with hard thresholding: among $n$ ratings, we incorporate only the ratings that are greater than or equal to a threshold $\tau$, denoted as \textit{sum}-$\tau$.

\paragraph{Reciprocal Rank Fusion (RRF).} 
Each multi-aspect rating can also be viewed as a separate score, resulting in multiple ranked lists (i.e., one list for each sub-question).
Under this view, a clear approach is to use reciprocal ranked fusion~\cite{cormack2009reciprocal}. The final score of the document $d$ is thereby obtained from $n$ distinct rankings with reciprocal rank normalization:
\begin{equation}
    s_{RRF}(d|x) = \sum_{j=1}^n \dfrac{1}{\kappa +{\rm Rank}_j(d)},
\end{equation}
where ${\rm Rank}_j(d)$ indicates different ranks of document $d\in \bar{C}$ sorted using the answerability of different sub-question $q_j$. 
Following common practice~\cite{cormack2009reciprocal}, we set $\kappa$ as 60.

\paragraph{Greedy Utility Selection (greedy-sum, greedy-$\alpha$, \& greedy-$cov$).}
Instead of naively aggregating multi-aspect ratings for each document independently, we adopt a greedy algorithm that iteratively selects the document maximizing some utility function (e.g., $\textit{sum}$, coverage, $\alpha$-nDCG).
The algorithm begins with an empty list $Z^{(0)}$. At each step $t$, we compute the utility of all the remaining document candidates $d\in\bar{C}$ and select the one that yields the highest utility gain. The selected one is then removed from the candidate set and appended to the list as $Z^{(t)}$.
Until the utility gains of every remaining document become zero, the remaining are then concatenated to the list in descending order of utility.

The utility function takes the document list $d\in Z^{(t)}$ as input. In our experiment, we implement several utility functions for a given document list $Z$. They are detailed as follows:
\begin{itemize}
\item \emph{greedy-sum} extends simple \emph{sum} aggregation in Eq.~\eqref{eq:sum} with first taking the maximum rating of each sub-question over documents in the list:
$$ U_{sum}(Z) = \sum_{j=1}^n \underset{d\in Z}{\max}\; r_{d, q_j}.$$
\item \emph{greedy-$\alpha$} is defined according to the denominator of evaluation metric $\alpha$-nDCG~\cite{Clarke2008-cn}. We first obtain the binary weight by applying a threshold $\tau$ on multi-aspect ratings. Then, we compute the ideal discounted cumulative gain with the penalty factor $\alpha$, which decays the gain given the counts of documents that covered the sub-question $q_j$, denoted as $c_{d_{i'<i}, j}$, formulated as:
$$ U_{\alpha}(Z) = \sum_{i=1}^{|Z|}\Big(\sum_{j=1}^n \mathbf{1} (r_{d_i, q_j}\geq \tau) \prod_{i'=1}^{i-1} (1-\alpha)^{c_{d_{i'}, j}} \Big).$$
\item \emph{greedy-$cov$} is derived from the coverage metric ($Cov$)~\cite{strecall,Ju2025-pn}, where we similarly take the maximum ratings of each sub-question over documents. Afterwards, we sum the binary weights to get the final coverage:
$$ U_{cov}(Z) = \sum_{j=1}^n \mathbf{1} (\underset{d\in Z}{\max}\; r_{d, q_j} \geq \tau).$$
\end{itemize}
These aggregation strategies enable LANCER to rank retrieved documents as a whole and to provide the refined retrieved context for the generator. Detailed parameter analysis is reported in Section~\ref{sec:ablation-agg}.

\section{Experiments and Results}

\subsection{Experimental Setup}\label{sec:exp-setup}
 \begin{table}[t]
    \centering
    \caption{Dataset statistics.}
    \setlength{\tabcolsep}{4pt}
    \resizebox{.7\textwidth}{!}{%
    \begin{tabular}{lccc}
    \toprule
         & NeuCLIR'24 ReportGen
         & CRUX-MDS DUC'04 \\
    \midrule
         \# Requests                        &  19            & 50      \\
         Avg. request length                &  55.95         & 48.46    \\
         Avg. \# nuggets per request        &  21.84         & 15      \\
         Corpus size                        &  10,038,768    & 565,015 \\
    \bottomrule
    \end{tabular}
     }
    \label{tab:data}
\end{table}

\subsubsection{Evaluation Datasets.} 
Evaluating information coverage requires nugget-level judgments, which are rare. We evaluate LANCER using two long-form RAG evaluation datasets: the TREC NeuCLIR'24 Report Generation (\textbf{NeuCLIR'24 ReportGen})~\cite{Dawn2025-sg}, and 
the CRUX multi-document summary with DUC'04 (\textbf{CRUX-MDS-DUC'04})~\cite{Ju2025-pn}.
Both evaluation datasets provide multi-faceted report requests along with corresponding nuggets, which indicate what information should be provided in the final generated report.
For NeuCLIR'24 ReportGen, we combine the 19 topics that were judged across all three languages as our test set and use the remaining 3 that are incomplete among the languages (topics 324, 361, and 387) as development topics. 
In NeuCLIR'24 ReportGen, each nugget is written in the form of a question with multiple acceptable answers. 
For CRUX-MDS-DUC'04, the nuggets are at the question-level and derived from a human-written summary. Dataset statistics are reported in Table~\ref{tab:data}.

\subsubsection{Evaluation Methods.}
To evaluate retrieval for long-form RAG, we adopt the CRUX framework~\cite{Ju2025-pn} to assess the quality of the intermediate retrieved context $Z$.
We report $\alpha$-nDCG and Coverage ($Cov$) for information coverage as our primary metrics. 
For reference, we also report
nDCG and precision (\text{Prec.}) to measure relevance.
All metrics are calculated with a rank cutoff at 10, given that a limited number of documents can typically fit in the input of the downstream generation models. 

\subsubsection{First-stage Retrieval.}\label{sec:exp-setup-inference}
In the following experiments, we adopt a standard two-stage retrieval pipeline to augment the final retrieved context for the report generation task, as formulated in Eq.~\eqref{eq:report-gen}.
First, we retrieve documents using one of three first-stage retrieval approaches:
BM25\footnote{The parameters $k_{1}, b$ are set to $(1.2, 0.75)$ for the NeuCLIR corpus; and $(0.9, 0.7)$ for the CRUX-MDS corpus.},
learned sparse retrieval (LSR) using MILCO~\cite{Nguyen2025-di} or SPLADEv3~\cite{Lassance2024-um}, and Qwen-3-Embed~\cite{qwen3embedding}.
NeuCLIR is a multilingual corpus; we use the official English translation of the corpus for BM25 and use documents in their source languages for LSR and Qwen3-Embed since they are natively multilingual models.
On NeuCLIR we use the MILCO multilingual LSR model~\cite{Nguyen2025-di}, whereas on DUC we use the English SPLADEv3 LSR model~\cite{Lassance2024-um}.
For each first-stage retrieval setting, we retrieve the top-$100$ candidate documents using the report request. 
These candidates are then passed to different second-stage reranking methods.

\subsubsection{Second-stage Reranking.}
\revised{We implement the other LLM reranking methods as comparable baselines, including 
\textbf{Pointwise}~\cite{Nogueira2019-kt, Zhuang2023-tl} with the document relevance estimated via softmax-normalized over ``Yes''/``No'' token logits,
\textbf{Listwise}~\cite{Sun2023-ln, Ma2023-gb} with a default window size of 20 and stride of 10, and
\textbf{Setwise} reranking~\cite{Zhuang2024-sx} with 5 child nodes and the heap sort algorithm.
All the reranking methods use \texttt{meta-llama/Llama-3.3-70B-Instruct}~\cite{Grattafiori2024-zw} with a suitable maximum context length.\footnote{10,240 for Setwise, 20,480 for Listwise, and 8196 for the others.} running on top of vLLM inference infrastructure.\footnote{https://github.com/vllm-project/vllm} 
The temperature is set to 0 for better reproducibility.
As a default, we generate 2 sub-questions\footnote{For CRUX-MDS-DUC'04, we use \texttt{Qwen/Qwen3-Next-80B-A3B-Instruct} for generating sub-questions, to avoid biases due to the fact that this dataset contains data synthesized by Llama 3.1~\cite{Ju2025-pn}.} and aggregate answerability ratings with $sum$ strategy. We explore the impact of these parameters in Section~\ref{sec:ablation}.}

\begin{table}[t]
    \centering
    \caption{
    Evaluation results on two datasets. 
    The first column group for each dataset contains relevance-based metrics, whereas the shaded columns report our primary coverage-based metrics. All metrics use a cut-off of 10. 
    Bold and underlined scores denote the \B{best} and \U{second-best} results within the same first-stage retrieval.
    Superscripts indicate when a metric shows a significant improvement over another approach according to a paired t-test as follows: first-stage($f$), Pointwise($p$), Listwise($l$), Setwise($s$), and LANCER($\dagger$).}
    \resizebox{\textwidth}{!}{%
    \setlength{\tabcolsep}{4pt}
    \begin{tabular}{l
    ll  >{\columncolor{gray!15}}l  >{\columncolor{gray!15}}l 
    ll  >{\columncolor{gray!15}}l  >{\columncolor{gray!15}}l 
    }
\toprule
    & \multicolumn{4}{c}{NeuCLIR'24 ReportGen} & \multicolumn{4}{c}{CRUX-MDS-DUC'04}  \\
    \cmidrule(lr){2-5}  \cmidrule(lr){6-9}  
    & \multicolumn{2}{c}{Relevance} & \multicolumn{2}{c}{\cellcolor{gray!15}\bf Coverage}
    & \multicolumn{2}{c}{Relevance} & \multicolumn{2}{c}{\cellcolor{gray!15}\bf Coverage} \\
    \cmidrule(lr){2-3}  \cmidrule(lr){4-5}  
    \cmidrule(lr){6-7}  \cmidrule(lr){8-9}  
    & nDCG & Prec. & $\alpha$-nDCG & $Cov$ & nDCG & Prec. & $\alpha$-nDCG & $Cov$ \\
\toprule
    BM25               & 67.7 & 65.3 & 53.0 & 64.1 & 53.0 & 51.4 & 44.5 & 54.1 \\
    + Pointwise        & \B{89.3} & \B{89.5} & \B{67.0} & \U{72.2} & 76.1 & 73.4 & 58.6 & 65.6 \\
    + Listwise         & 86.0 & 84.7 & 61.3 & 69.5 & \B{77.1} & \B{74.6} & \U{60.3} & 64.9 \\
    + Setwise          & 84.2 & 81.6 & 64.5 & 71.3 & 69.2 & 64.0 & 57.8 & 63.5 \\
    + LANCER           & 86.2$^{f}$ & 85.8$^{f}$ & \U{65.5}$^{fl}$ & \B{72.7}$^{fls}$ 
                       & 73.8$^{fs}$ & 72.4$^{fs}$ & \B{60.5}$^{fs}$ & \B{66.4}$^{fs}$ \\
    \hdashline \noalign{\vskip 1pt}
    + LANCER$_{Q^*}$   & 88.0$^{f}$ & 85.8$^{f}$ & 76.7$^{fpls\lc}$ & 79.1$^{fpls}$ 
                       & 80.3$^{fpls\lc}$ & 76.6$^{fps\lc}$ & 73.7$^{fpls\lc}$ & 74.6$^{fpls\lc}$ \\
\midrule
    LSR                & 83.1 & 81.6 & 62.9 & 73.7 & 70.4 & 68.0 & 55.8 & 64.0 \\
    + Pointwise        & 90.7 & 90.5 & 66.4 & 72.9 & \B{83.1} & \B{82.4} & \U{63.2} & \B{70.6} \\
    + Listwise         & 92.3 & 90.5 & \U{71.2} & \U{74.9} & 80.6 & 79.8 & 61.1 & 67.1 \\
    + Setwise          & 91.0 & 89.5 & 68.6 & 72.1 & 71.3 & 68.4 & 58.4 & 64.8 \\
    + LANCER           & \B{92.9}$^{f}$ & \B{91.6}$^{f}$ & \B{72.4}$^{fp}$ & \B{77.3}
                       & 78.9$^{fs}$ & 78.0$^{fs}$ & \B{63.5}$^{fs}$ & \U{68.8}$^{fls}$ \\
    \hdashline \noalign{\vskip 1pt}
    + LANCER$_{Q^*}$   & 90.9$^{f}$ & 90.5$^{f}$ & 78.9$^{fpls\lc}$ & 81.4$^{fpls\lc}$ 
                       & 86.3$^{fpls\lc}$ & 84.8$^{fls\lc}$ & 81.3$^{fpls\lc}$ & 79.3$^{fpls\lc}$ \\
\midrule
    Qwen3-Embed        & \B{88.6} & \B{86.8} & 62.7 & 69.5 & 75.9 & 73.8 & 60.8 & 66.8 \\
    + Pointwise        & 85.1 & 86.3 & 63.3 & 69.6 & \B{83.9} & \B{83.8} & 63.0 & \B{70.2} \\
    + Listwise         & 88.4 & \B{86.8} & 65.1 & 68.4 & 81.7 & 81.4 & \U{63.2} & 67.5 \\
    + Setwise          & 84.6 & 80.5 & \U{68.5} & \U{71.2} & 75.3 & 72.4 & 61.9 & 66.6 \\
    + LANCER           & 88.0 & 85.3 & \B{70.7}$^{fpl}$ & \B{75.3}$^{fp}$ 
                       & 80.8$^{fs}$ & 80.0$^{fs}$ & \B{64.4}$^{fs}$ & \U{68.5}$^{fs}$ \\
    \hdashline \noalign{\vskip 1pt}
    + LANCER$_{Q^*}$   & 88.6 & 88.4$^s$ & 78.8$^{fpls\lc}$ & 80.8$^{fpls\lc}$ 
                       & 88.2$^{fpls\lc}$ & 86.8$^{fpls\lc}$ & 82.9$^{fpls\lc}$ & 80.1$^{fpls\lc}$ \\
\bottomrule
    \end{tabular}
    }
    \label{tab:main-1}
\end{table}

\subsection{Main Results}\label{sec:main}
Table~\ref{tab:main-1} presents our empirical evaluation results on NeuCLIR'24 ReportGen and CRUX MDS-DUC'04, where each block of rows corresponds to distinct first-stage retrievers.
In addition to the targeted coverage-based metrics ($\alpha$-nDCG@10 and $Cov$@10, shown in the last two shaded columns), we also report relevance-based metrics for reference (the first two columns).

\subsubsection{Zero-shot Reranking Comparisons.}
We compare our proposed LANCER to three common LLM-based reranking methods:  
\textbf{Pointwise}~\cite{Nogueira2019-kt, Zhuang2023-tl}, \textbf{Listwise}~\cite{Sun2023-ln, Ma2023-gb}, and 
\textbf{Setwise} reranking~\cite{Zhuang2024-sx}.
Improvements are observable across three different first-stage retrieved candidate environments (different blocks) and both evaluation datasets, showing the reranking robustness and generalizability.
However, we found that the improvements are relatively minor on CRUX-MDS-DUC'04 in terms of $Cov$, where we attribute this to the smaller number of ground-truth nuggets, limiting the possible number of documents that can be credited in $Cov$, and, thus, lower scores. 
These improvements are sometimes significant, but it is difficult to reach statistical significance given the small sizes of the available datasets with nugget-level judgments. 

\subsubsection{Trade-off Between Relevance and Coverage.}
In addition, we observe trade-offs between reranking for relevance and coverage.
Relevance-based reranking improves first-stage retrieval in terms of nDCG and precision; however, their gains on the coverage-based metrics are limited and even slightly decreased when using stronger first-stage retrievers.
For example, LSR with Setwise reranking increases nDCG and precision but reduces coverage (-2.8), suggesting that reranking for relevance can filter out irrelevant documents but may fail to pull up documents that cover different aspects. 
On the contrary, LANCER achieves better coverage without trading off much relevance.  
Using Qwen3-Embed on NeuCLIR'24 ReportGen, LANCER outperforms Listwise reranking on Coverage (75.3 vs. 68.4) but without substantially reducing precision (85.3 vs. 86.8). 
Listwise reranking with Qwen3-Embed, while achieving better precision (only higher than LANCER by 1.5 points), is 5 points lower in $\alpha$-nDCG and almost 7 points lower in Coverage.
On NeuCLIR ReportGen with LSR as the first stage, LANCER even exhibits stronger performance on both relevance and coverage effectiveness than the baselines, showing no trade-off between the two.

\subsubsection{Oracle Setting with Ground-truth Sub-questions.}
To explore the optimal effectiveness of LANCER, we replace the $n$ synthetic sub-questions with ground-truth nugget questions to remove the noise of sub-question generation and control other inference settings, 
including the LLM generation and the ranking optimization strategy as \emph{sum}. 
This oracle condition is denoted as LANCER$_{Q^*}$ and reported in the last row of each block in Table~\ref{tab:main-1}.
With the ground-truth nugget questions, LANCER achieves substantially higher $\alpha$-nDCG and $Cov$ compared to the other methods. 
\revised{This illustrates that the sub-questions are crucial for optimizing nugget-coverage, echoing previous work on the challenging research directions of nugget generation for RAG~\cite{Lajewska2025-wv, Pradeep2024-fe} and highlighting nugget generation as a promising direction for improvement.}

\begin{figure}[t]
    \centering
    \includegraphics[width=\textwidth]{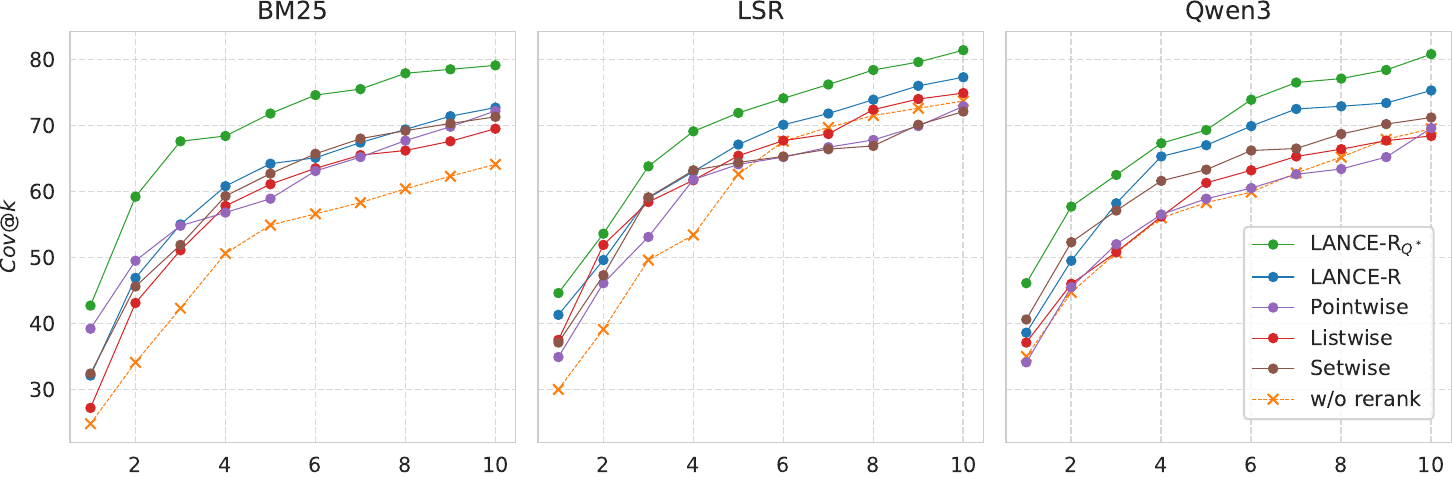}
    \caption{
        Coverage ($Cov$) grows with respect to the top-$k$ cutoff on NeuCLIR'24 ReportGen evaluation data. Each line indicates the retrieved contexts from different retrieval pipelines.}
    \label{fig:coverage-at-k}
\end{figure}

\subsubsection{Top-ranking Retrieved Context.}
We further analyze coverage with different cutoffs (top-$k$) from LANCER.
Figure~\ref{fig:coverage-at-k} shows Coverage@10 at different depths, reporting the dynamics of how the multi-aspect information needs are satisfied.
Notably, with LSR and Qwen3-Embed as first-stage retrieval, we found that, after $k=4$, LANCER starts to outperform the other reranking approaches and consistently improves as $k$ increases.
The trend is less consistent with BM25, though LANCER with oracle nugget questions still eventually achieves similar Coverage with larger $k$.
This difference may be due to the fact that the retrieved candidates were  harder for any reranking methods to distinguish due to lexical overlap with the original queries~\cite{Alaofi2025-qd}. 
\revised{Nevertheless, LANCER performs well overall and LANCER with oracle nuggets still consistently outperforms other methods across different top-$k$ and first-stage retrieval. We therefore set a future goal of conducting an in-depth investigation into reducing the gap between synthesized and ground-truth sub-questions.}

\subsubsection{Impact of Retrieved Context on Generation.}
To analyze the downstream impact of LANCER on the generated report, we additionally measure the nugget-coverage of the final RAG result $y$.
We employ \texttt{GPTResearcher}~\cite{duh2025hltcoeliveraggptresearcherusing},\footnote{\url{https://github.com/assafelovic/gpt-researcher}} an open-source report generation method, and input it with the different retrieved contexts to produce final reports.
The report nugget-coverage scores are obtained from Auto-ARGUE~\cite{William2025-tq}, an automatic evaluation framework implementing ARGUE~\cite{Mayfield2024-vw} with \texttt{Llama-3.3-70B-Instruct}.
For a fair comparison, we fix the generation settings and use the same number of top-$k$ retrieved documents\footnote{$k$ is set to 8 and the temperature is set to 0.35.} for all 18 retrieval pipelines (rows in Table~\ref{tab:main-1}).
We observe a good Spearman correlation (0.78 and 0.7) between the report nugget-coverage (percentage of the unique gold nuggets in the generated report) and the two coverage-based evaluation metrics ($\alpha$-nDCG and $Cov$, respectively).
Notably, LANCER with oracle nugget questions achieves additional +3.5 and +4 nugget-coverage over other reranking methods. However, it only changes +1.3 when paired with LSR first-stage retrieval, which may indicate noise in the downstream generation steps in incorporating information in retrieved documents
due to various known issues such as positions~\cite{Liu2024-zv}, content~\cite{belem2025single}, and parametric memory~\cite{xie2024adaptive}.

\subsection{Parameter Analysis}\label{sec:ablation}

\begin{figure}[t]
    \centering
    \includegraphics[width=.8\linewidth]{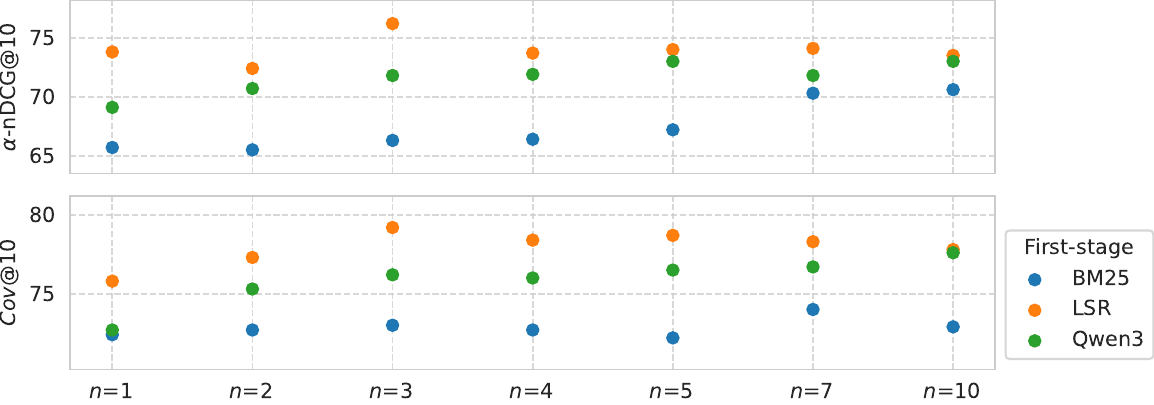}
    \caption{
    Evaluation results on the NeuCLIR’24 ReportGen with different numbers of synthetic sub-questions ($x$-axis). We use the \emph{sum} strategy for all the settings. The colors indicate the three first-stage retrieval.
    }
    \label{fig:ablation-nq}
\end{figure}

\subsubsection{Number of Sub-questions.}\label{sec:ablation-nq}
Figure~\ref{fig:ablation-nq} shows how varying the number of synthetic sub-questions in the LANCER pipeline affects $\alpha$-nDCG and $Cov$ on NeuCLIR ReportGen.
Surprisingly, the results suggest that a few sub-questions (2 or 3) are sufficient, while adding more does not substantially reduce performance  compared to $n=2$ but only offers a marginal benefit. 
When using BM25, increasing the number of sub-questions yields a more substantial improvement compared to the other first-stage retrievers, with $\alpha$-nDCG@10 increasing as $n$ increases and the highest \textit{Cov} at $n=7$.
Diminishing returns and drops in performance may be due to topic drift as the number of sub-questions increases.
However, this is not the case for the oracle nugget questions, which are more than 10 but contribute significant benefits on coverage, indicating more sub-questions can still be useful if they remain aligned with the original information need. 
We leave such question generation for future work aiming to explore more useful questions for LANCER.

\begin{figure}[t]
    \centering
    \includegraphics[width=\linewidth]{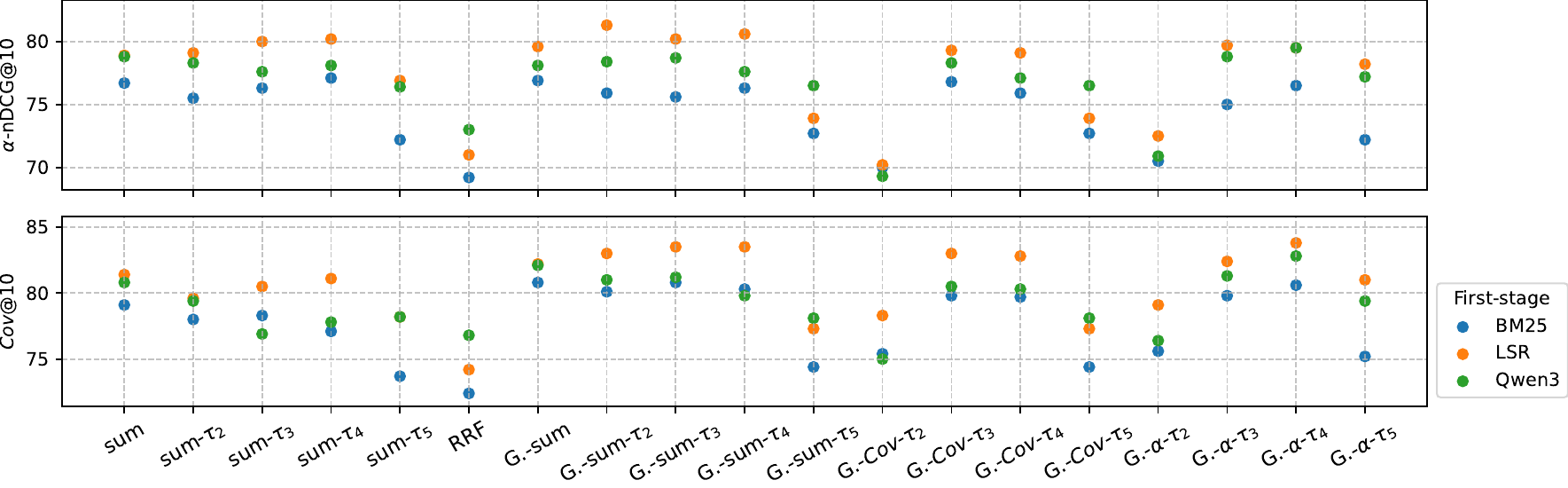}
    \caption{
    Evaluation results on the NeuCLIR’24 ReportGen. The $x$-axis shows different aggregation strategies. The colors indicate the three first-stage retrieval.
    }
    \label{fig:ablation-agg}
\end{figure}

\subsubsection{Different Optimization Strategies.}\label{sec:ablation-agg}
In addition, we investigate different strategies of utilizing multi-aspect ratings $r_{d, q_j}$.
To control the impact of synthetic questions, we adopt the oracle nugget questions to judge answerability (i.e., LANCER$_{Q^*}$).
Figure~\ref{fig:ablation-agg} shows the results of applying 5 strategies (Section~\ref{sec:strategy}) with or without thresholding $\tau\in[2,5]$.
We found that the \emph{sum} strategy generally performs well on $\alpha$-nDCG. In contrast, greedy selections (\emph{G.-*}) achieve better $Cov$ at threshold 3 or 4 ($\tau_3, \tau_4$), which is what it is optimizing for. 
However, interestingly, they drop substantially when applying thresholds at 2 or 5. 
An exception is \emph{greedy-sum}, which combines ratings additively, and thus is less sensitive to thresholding. 
\revised{These empirical results imply that the human's nugget identification aligns closer to an LLM answerability judgment of 3 or 4, presenting the fact that there is an uncertainty of LLM-judgment especially when the predicted rating is low.}
\revised{To effectively reduce noise and integrate lower ratings better in LANCER, we hypothesize incorporating the logit-trick~\cite{Nogueira2020-vn, Gangi-Reddy2024-vk} has potential to address this issue, as evidenced in~\citet{Zhuang2023-bs} and by the observed performance of Pointwise reranking in Table~\ref{tab:main-1}.}

\section{Conclusion}
In this paper, we propose LANCER, an LLM re-ranking method that targets nugget-coverage for the retrieval of long-form RAG. As opposed to existing relevance-based retrieval approaches, LANCER generates sub-questions as proxy nuggets and produces multi-aspect ratings with a coverage-based aggregation. 
\revised{Empirical evaluation shows that LANCER is able to effectively rank documents based on nugget coverage without losing the ability to perform relevance ranking, highlighting its suitability as a retrieval method for long-form RAG tasks.}
\revised{Our analyses further highlights promising directions for future nugget-coverage optimization, as evidenced by the quality of proxy sub-question and unstable LLM answerability judgment.}

\section*{Acknowledgments}
This research was supported by the \href{https://hybrid-intelligence-centre.nl}{Hybrid Intelligence Center}, a 10-year program funded by the Dutch Ministry of Education, Culture and Science through the Netherlands Organisation for Scientific Research, project VI.Vidi.223.166 of the NWO Talent Programme which is (partly) financed by the Dutch Research Council (NWO) and NWO project NWA.1389.20.\-183.
We acknowledge the Dutch Research Council for awarding this project access to the LUMI supercomputer, owned by the EuroHPC Joint Undertaking, hosted by CSC (Finland) and the LUMI consortium through project number NWO-2024.050.
Views and opinions expressed are those of the author(s) only and do not necessarily reflect those of their respective employers, funders and/or granting authorities.

\section*{Disclosure of Interests}
The authors have no competing interests to declare that are relevant to the content of this article.



\bibliography{references}

\end{document}